\documentclass[]{article}
\usepackage{emulateapj}
\usepackage{graphicx}

\def\msun{{\rm ~M}_{\odot}}

\def\mdot{\dot M}

\begin{document}

\title{The Nature of the Faint {\em Chandra} X-ray Sources in the Galactic Centre}

 \author{Ashley J. Ruiter\altaffilmark{1}, Krzysztof Belczynski\altaffilmark{1,2}, 
         and Thomas E. Harrison\altaffilmark{1}}

 \affil{
     $^{1}$ New Mexico State University, Dept. of Astronomy,
        1320 Frenger Mall, Las Cruces, NM 88003\\
     $^{2}$ Tombaugh Fellow\\
     aruiter,kbelczyn,tharriso@nmsu.edu}

 \begin{abstract} 
Recent {\em Chandra} observations have revealed a large population of faint 
X-ray point sources in the Galactic Centre. The observed population consists 
of $\gtrsim 2000$ faint sources in the luminosity range $\sim$ $10^{31}$-$10^{33}$ 
erg s$^{-1}$. The majority of these sources (70\%) are described by hard spectra, 
while the rest are rather soft. The nature of these sources still remains unknown. 
Belczynski \& Taam (2004) demonstrated that X-ray binaries with neutron star or 
black hole accretors may account for most of the soft sources, but 
are not numerous enough to account for the observed number and X-ray 
properties of the faint hard sources. 
A population synthesis calculation of the Galactic Centre region has 
been carried out. Our results indicate that the numbers and X-ray luminosities of 
intermediate polars are consistent with the observed faint hard Galactic Centre 
population. 
 \end{abstract}


\section{Introduction}

A {\em Chandra} X-ray survey of the Galactic Centre (GC, Wang, Gotthelf \& Lang 
2002) first revealed the presence of $\sim$ 1000 spectrally hard X-ray sources 
(2-10 keV) with luminosities $L_{\rm x}
\lesssim 10^{35}$ erg s$^{-1}$. Pfahl et al. (2002) have claimed that wind-fed 
neutron stars (NS) with intermediate- and high-mass companions are 
responsible for a significant fraction of the hard sources in the Wang et al. 
(2002) survey. 
A deeper {\em Chandra} survey of the nuclear bulge region (Muno et al. 2003;
hereafter MM03) 
revealed over 2000 X-ray point sources with X-ray luminosities $\sim$ $10^{31}$
-$10^{33}$ erg s$^{-1}$. The majority of these sources (1427)
are described by hard spectra (with a photon index of an absorbed power law
$\Gamma < 1$), while the rest (652) are characterized by softer spectra 
($\Gamma > 1$).\footnote{Foreground sources excluded; M.Muno, private
communication.}
Belczynski \& Taam (2004; hereafter BT04) have studied the entire population of 
X-ray binaries with NS and black hole (BH) accretors in the context of the MM03 
survey. It was demonstrated that neither wind-fed (low-, intermediate-, 
high-mass) systems nor Roche Lobe Overflow (RLOF) binaries can explain 
the entire faint population. However, the quiescent transients may be responsible 
for most of the faint soft GC sources. 
Muno et al. (2004) suggested that the observed faint hard sources are most likely
intermediate polars (IPs); a subclass of magnetic cataclysmic variables (CVs).

Among magnetic CVs, IPs are asynchronous rotators, with the white 
dwarf (WD) spin period being shorter than the orbital period. Currently there 
are $\sim$ 31 IPs known (Gansicke et al. 2005). They consist of a 
magnetized WD and a low-mass companion. The companion transfers mass via RLOF to the 
WD, with typical mass transfer (MT) rates of $\sim 10^{-11}$ M$_{\rm \odot}$/yr.
Matter spirals in toward the WD forming an accretion disc, 
which is truncated by the WD magnetosphere.  Matter is then channeled into
accretion columns over both magnetic poles of the WD (e.g., Belle et al. 2005). 
Typical IP magnetic fields are $\leq$ 10 MG (de Martino et al. 2004). It is
from below the accretion shock, above the WD surface, where the hard X-rays
originate (see Warner 1995; Patterson 1994 for review).
IPs are known to exhibit both soft and hard X-ray emission and are thought 
to be the most luminous subclass of CVs in X-rays, owing to their typically 
higher MT rates.    

We address the issue of the nature of the faint hard source population observed in 
the deep GC exposure of MM03, and test the validity of the Muno et al. 
(2004) hypothesis that these sources are IPs. We construct a simple phenomenological 
model of an IP and use population synthesis (\S\,2) to calculate the number of IPs 
and their X-ray luminosities in the GC (\S\,3). In \S\,4 we comment on other
population synthesis studies and discuss our results in context of available 
observations. 

\section{Model Description}

We use the updated population synthesis code
{\tt StarTrack} (Belczynski, Kalogera \& Bulik 2002; Belczynski et al. 2005a).   
All stars are evolved with solar metallicity ($Z$=0.02) and with standard wind 
mass loss rates. We assume a continuous star formation rate in the GC over the last 
10 Gyrs, and a binary fraction of 50\%.  
Several physical processes important for binary evolution are accounted for: 
tidal interactions, detailed MT calculations, 
common envelope (CE) events, supernovae explosions, and various mechanisms
for angular momentum losses such as magnetic braking (MB) and gravitational 
radiation (GR), among others. 
 
The goal of this study is to test the hypothesis that IPs can account 
for the observed population of the faint hard X-ray point
sources in the GC.  We use our standard model (Belczynski et al. 
2005a) to check whether within the general framework of binary evolution, the 
number of predicted IPs coincides with that of the GC faint hard sources. 
It has been realized that CVs are the likely 
outcome of the CE phase (Paczynski, 1976).  Despite years of 
investigation, no consistent physical model for this phase of binary 
evolution exists.  Therefore, we use two current alternative pictures of 
CE evolution to provide an estimate of associated model uncertainties.  
Our standard CE model calculations assume energy balance (Webbink 1984); the
donor envelope is ejected from the system at the expense of the binary orbital 
energy. 
The binary separation following the CE phase depends on parameters $\alpha_{\rm ce}$ 
and $\lambda$, relating to the efficiency with which the orbital energy is
used to expel the CE from the 
system, and donor internal structure, respectively. We adopt $\alpha_{\rm ce} \times 
\lambda = 1.0$. 
In our alternative CE model, we have adopted the prescription of Nelemans \& Tout (2005) 
which employs angular momentum balance, and assumes 
that the angular momentum is lost from the binary in a linear fashion as a function 
of mass loss. Following Nelemans \& Tout (2005), we adopt a scaling factor 
$\gamma = 1.5$ for this model. See Belczynski, Bulik \& Ruiter (2005b) for the CE equations.

{\em X-ray Luminosity Calculation.}
We assume that the IP X-ray luminosity is a function of the accretion rate,
accretor physical properties, and the efficiency with which the accretion 
luminosity is converted to hard X-ray luminosity in the {\em Chandra} band.

For degenerate donors ($M_{\rm don}$), we assume that MT is GR-driven and 
calculate it via  
\begin{equation}
\mdot_{\rm don} = M_{\rm don} D^{-1} {\,d J_{\rm gr}/\,d t \over J_{\rm orb}}
\label{mt13}
\end{equation}
\begin{equation}
D={5 \over 6}+ {1 \over 2} \zeta_{\rm don}-{1-f_{\rm a} \over 3 (1+q)}-
{ (1-f_{\rm a}) (1+q) \beta_{\rm mt}+f_{\rm a} \over q}
\label{mt17}
\end{equation}
where $J_{\rm orb}$ is the orbital angular momentum, $dJ_{\rm gr}/dt$ is the angular 
momentum loss due to GR, $f_{\rm a}$ is the fraction of transferred mass 
accreted by the WD of mass $M_{\rm acc}$ (we assume Eddington limited 
accretion; see below), $q \equiv (M_{\rm acc}/M_{\rm don})$, and $\beta_{\rm mt} 
= M_{\rm acc} M_{\rm don}^{2}/(M_{\rm don} + M_{\rm acc})^{2}$. 
The radius mass exponent for the donor $\zeta_{\rm don}$ is 
obtained from stellar models (see Belczynski et al. 2005a).  

For non-degenerate donors we estimate the MT rate as follows: 
\begin{equation}
\mdot_{\rm don} =  - {\zeta_{\rm evl}+{2 \over \tau_{\rm mb}} + {2 \over
                    \tau_{\rm tid}} + {2 \over \tau_{\rm gr}}  \over
                    \zeta_{\rm don} - \zeta_{\rm lob}} M_{\rm don}
\label{mt7}
\end{equation}
where $\zeta_{\rm evl}$ is the change of the donor radius due to its nuclear 
evolution, $\zeta_{\rm lob}$ is the radius mass exponent for the donor Roche lobe, 
and $\tau_{\rm mb}$, $\tau_{\rm tid}$ and $\tau_{\rm gr}$ are the timescales 
associated with MB, tidal interactions and GR, respectively.
In some cases MT proceeds on a thermal timescale, and thus we use $\mdot_{\rm th} 
= - {M_{\rm don}\tau_{\rm th}}$
where the thermal timescale may be obtained from $\tau_{\rm th} = ({30 \times
{M_{\rm don}^{2}}) / (R_{\rm don} L_{\rm don}})$ (e.g., Kalogera \& Webbink
1996).             
Then, we can calculate mass accretion rate as 
$\mdot_{\rm acc}={\rm min}(0.95 \times \mdot_{\rm don}, \mdot_{\rm edd})$. 
We limit the accretion rate to the critical Eddington rate
($\mdot_{\rm edd}$), such that if MT is super-Eddington, the 
excess material leaves the binary with the specific angular momentum 
of the accretor. We impose a mass-loss rate of 5\% 
in the sub-Eddington regime since some IPs are observed to experience a mass 
loss rate of few percent ($\sim 10$\% in the extreme case of AE Aqr;
Wynn, King \& Horne 1997). 

The hard {\em Chandra} band X-ray luminosity of an IP system is calculated 
from 
\begin{equation} 
L_{\rm x} = \eta_{\rm bol}\, \eta_{\rm geo} \, L_{\rm bol} = \eta_{\rm bol} \, 
\eta_{\rm geo}\, \epsilon \, {G M_{\rm acc} \mdot_{\rm acc} \over R_{\rm acc}}
\label{Lx}
\end{equation} 
where $G$ is the gravitational constant, $\epsilon$ is the conversion efficiency 
of gravitational binding energy to radiation (1 for WD surface accretion), and 
$M_{\rm acc}$ and $R_{\rm acc}$ are the accreting WD mass and radius, 
respectively.  We use $\eta_{\rm bol}=0.09$ for 
the bolometric correction to 2-8 keV X-ray luminosity, although it is noted 
that this value is quite uncertain and may span a wide range ($\sim$
0.01-0.2; Norton \& Watson 1989b). 
The IP X-ray emission is likely anisotropic to some extent, since
the accretion proceeds through the WD magnetic poles. However, it has been  
demonstrated that the emission region in IPs may be quite extensive (i.e., may 
encompass more than a quarter of the WD surface, Norton \& Watson 1989a) 
and thus the anisotropy is not expected to be large.
Here we assume isotropic emission and so $\eta_{\rm geo}=1$, although we
note that a more sophisticated model including polar cap accretion should
be invoked once more observations are available. In general $\eta_{\rm
geo}$ depends on the size of the emitting regions and their relative
orientation to the observer.

{\em Initial Conditions.}
The initial mass of single stars and the primary (more massive) components in
binaries are drawn within the range 0.8-150 $\msun$ from  a broken 
three-component power law initial mass function (IMF), with a slope of
-1.3/-2.2/-2.7 in mass ranges 0.08-0.5/0.5-1/1-150 $\msun$ (Kroupa, Tout
\& Gilmore 1993). The secondary mass is obtained through a flat mass ratio
distribution. Initial binary orbits are specified by semi-major axis 
(distribution flat in the logarithm $\sim 1/a$) and eccentricity (thermal 
distribution $\sim 2e$). 

The stellar population of the GC is evolved through 10 Gyr (age of the 
Galaxy) with a constant star formation rate and IP systems are 
extracted and luminosities compared against the MM03 survey sources. We call 
{\em any} binary system experiencing RLOF in which the accretor is a
WD and the donor is any type of star a CV (i.e. including AM CVn systems, 
see Warner 1995). We consider various types of WDs: helium (HeWD), carbon-oxygen 
(COWD), oxygen-neon (ONeWD), hydrogen (HWD; formed by stripping the envelope of 
a low-mass main sequence (MS) star in RLOF), and hybrid (HybWD; containing a
He envelope and a He-C-O mantle, formed by stripping of a low-mass He star 
in RLOF). We note that the evolution of IPs and CVs are identical in 
our model, i.e., we do not incorporate the magnetic field of the WD into the 
evolution.  Rather, from the CV population we assign a 
fraction of CV systems that are IPs. In our model we adopt 
an IP fraction (IP$_{\rm frac}$) of 5\% following MM03. 
We have calibrated our results to pertain to the surveyed GC region by scaling by 
stellar mass. Following Muno et al. (2004; see their \S\,3), we have assumed that 
the 17'$\times$17' {\em Chandra} field of view of the GC corresponds to a cylinder 
440 pc deep with a radius of 20 pc, encompassing $1.3 \times 10^{8}$ M$_{\odot}$ 
in stars\footnote{There may be 50\% uncertainty associated with this
estimate (Launhardt et al. 2002), which will propagate linearly in our
results.}. 

\section{Results} 

\begin{figure}
\includegraphics[width=0.6\columnwidth,angle=0]{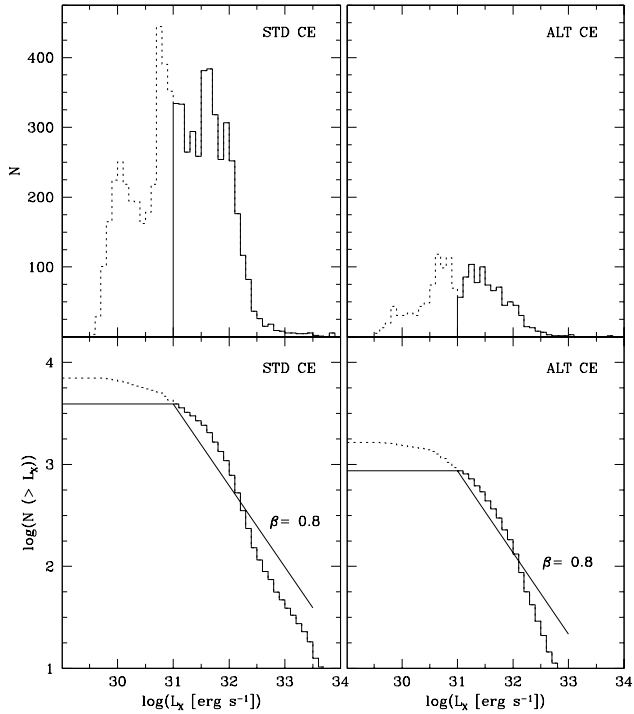}
\caption{Left: X-ray luminosity distribution and corresponding XLF for 
GC IPs for the standard CE model and an IP fraction of 5\%. The entire 
population is shown with a dotted line while IPs brighter than the MM03 
X-ray luminosity threshold are shown with a solid line.  
The XLF slope of the cumulative distribution ($N(>L_{\rm x}) \sim L_{\rm x}^{-\beta}$) 
is marked. Right: same as the left panels but for the alternative CE model.
}
\label{fig01}
\end{figure}

It is found that the synthetic GC population of IP systems span a 
range of X-ray luminosities $\sim 3 \times 10^{29} - 5 \times 10^{33}$ 
erg s$^{-1}$. In the following we only discuss systems above the MM03 survey 
detection limit, i.e., with X-ray luminosities $\geq 10^{31}$ erg 
s$^{-1}$ unless otherwise noted. The main results of our calculations for 
the standard and alternative CE prescriptions are presented in Table~\ref{tab01} 
and Figure~\ref{fig01}. 

{\em Standard CE model.}
The number of IPs depends strongly on the adopted IP$_{\rm frac}$. We find
$\sim$ 800-8000 IPs in the GC for IP$_{\rm frac}$ of 
1-10\%, respectively. To match the observed number of faint hard X-ray sources in the 
GC (1427) with IPs we would require an IP$_{\rm frac}$ of $\sim 2\%$.
The distribution of orbital periods of our IPs peaks between $\sim$ 1.5-3.5 
hr and extends toward longer periods ($\geq 5$ hr), which is in agreement 
with the observed orbital periods of magnetic CVs (see i.e., Tovmassian et al. 2004). 
We find that the most frequent IP type (89\%) is a magnetic WD accreting 
from a MS star.  The majority of donors (87\%) are $\le 0.7$ M$_{\odot}$,  
i.e. K-M dwarfs. Accretors in the WD-MS subclass are COWDs (66\%) and HeWDs 
(23\%). The second most frequent (10.5\%) type of IP is a double degenerate system 
(i.e., close WD-WD binary with stable RLOF). The most common configurations found
within this subclass are: COWD-HeWD (6.6\%), COWD-HybWD (1.7\%) and COWD-HWD (1.7\%).  
Finally, we find few (0.5\%) IPs with giant-like donors. 
In this subclass nearly all systems consist of a low-mass 
($\sim 0.3-0.4 \msun$) sub-giant or a red giant transferring mass to a COWD. 
The relative occurrence frequencies of different IP types reflect  
various evolutionary timescales for donors of varying types and masses.
The WD-MS IPs are abundant since low-mass MS stars have very long evolutionary 
timescales ($\sim 10^{11}$-$10^{12}$ yr) and RLOF at the IP phase is also driven 
on long timescales; by GR ($\tau_{\rm gr} \sim 10^{10}$ yr) with the addition of 
MB ($\tau_{\rm mb} \gtrsim 10^9-10^{10}$ yr) for some systems. 
The evolution of double degenerates is driven by GR on shorter timescales on  
the order $\tau_{\rm gr} \sim 10^9$ yr (since these binaries are tighter); 
moreover it is difficult to form these systems (survival of two CE phases rather than one). 
For WD-giant donor systems, the evolutionary and MB timescales are shorter 
than in the above cases, making them the least represented subclass of IPs.  

In Figure~\ref{fig01} we show the overall luminosity distribution 
and the corresponding X-ray luminosity function (XLF) for the model IPs. 
The IPs above the MM03 detection limit are marked for
easy comparison. It is noted that only $\sim 55$\% of IPs are bright enough 
to make the X-ray luminosity threshold of the survey. The brightest IPs in our
simulations are those with giant donors, with average X-ray luminosities 
$L_{\rm x} \sim 5 \times 10^{32}$ erg s$^{-1}$. 
The power-law slope of the cumulative distribution ($N(>L_{\rm x}) \sim 
L_{\rm x}^{-\beta}$) of the XLF for the synthetic IPs (above the survey detection 
limit) is $\beta \sim 0.8$. This value is shallower than the one derived by
MM03\footnote{Although formula (5) in MM03 does not represent a cumulative 
distribution, the listed value of the slope is cumulative; M.Muno private communication.}
($1.34$ for $L_{\rm x} > 4 \times 10^{31}$ erg s$^ {-1}$) and Muno et al. (2006).
It is noted that the full XLF slope modeling for IPs would require {\em i})
taking into account absorption toward the GC sources (as well as circumbinary 
absorption), and {\em ii}) a detailed model of anisotropic X-ray emission from a 
magnetized WD. 

{\em Alternative CE model.}
It is found that $\sim$ 170-1700 IPs may be present in the GC for IP$_{\rm
frac}$ of 1-10\%, and one would require an IP$_{\rm frac}$ of $\sim 8\%$ to
match the number of GC faint hard sources. Orbital periods are found in the
same range as for the standard CE model.
Once again, the most frequent IP type is a magnetic WD accreting from a
low-mass MS star (90.8\%) with the majority of them being COWD-MS star binaries (75\%).
Double degenerates are the second most populated subclass (8.6\%),
first with COWD-HWD (4\%), followed by COWD-HeWD and then COWD-HybWD (2.1\% 
and 1.9\%, respectively) with other types constituting the rest. 
Again only a small fraction (0.6\%) of the IPs involve a WD-giant star 
binary.
The slope of the XLF for IPs above the survey detection limit is found again to 
be $\beta \sim 0.8$ (see Fig 1).  

Additionally, we have calculated a model with decreased CE efficiency 
($\alpha_{\rm ce} \times \lambda = 0.1$). The number of IPs decreases by
$\sim$ 50\%, due to the more frequent binary mergers during CE phases.
Also, we have tested the assumption on the initial mass ratio of binary
components with a model in which both binary component masses are chosen 
independently from the IMF. The overall binary evolution is quite different, 
since on average $q$ is much smaller ($\sim$ 0.1, as opposed to the flat
$q$-distribution in other models). There is an overall increase ($\sim$ 20\%) 
in the number of IPs, though the population content is very similar with 
89\% of IPs with $> L_{\rm x} 10^{31}$ erg s$^{-1}$ being MS-WD binaries.        

{\em Typical evolution.}
Two MS stars (2 and 0.6 $\msun$) start out on a highly eccentric 
orbit ($e$=0.9) with an orbital period of 1460 days when the Galaxy is 2.1 
Gyr old. At 3.6 Gyr the system has circularized (period of 109 days), a CE phase 
is initiated by the primary (now an asymptotic giant), and upon ejection of the 
envelope the orbital period shrinks by $\sim$ two orders of magnitude (now 16 hours). 
The primary shortly thereafter becomes a COWD. At 8.4 Gyr the MS  
secondary initiates RLOF and the system becomes an IP. At 10 Gyr we find an IP 
system with a period of 2.9 hours.  
We note that for this same system, if the alternative CE prescription model is 
used an IP is never formed.  The reason for this is that upon ejection of the 
CE, the binary does not lose enough orbital angular momentum in order to end up 
in a tight enough orbit such that RLOF, and an IP phase, can ensue. Hence a 
smaller number of IPs are found in our alternative CE model.  

\section{Discussion}

We have explored the possibility that the faint hard sources in the GC are
IPs. It is found that for both current CE models, the IP population is ample 
enough to explain the GC faint hard sources. The required IP fractions are then 
$\sim 2\%$ and $\sim 8\%$ for the standard and alternative CE models, respectively. 
The actual fraction is uncertain and currently estimated to be $\sim 5\%$ by 
MM03 based on the Kube et al. (2003) catalog of CVs. Once more stringent observational 
constraints on the IP fraction are obtained, we will be able to test the validity of
different CE models. 

At this point a full picture of the GC X-ray point sources begins to emerge.  
Most of these sources are faint and are spectrally hard (1427) 
and can be explained by a population of IPs as 
suggested by Muno et al.(2004) and confirmed in this study.  Additionally, there may 
be a small contribution ($\sim$ few percent) of wind-fed sources 
with NS and BH accretors to the faint hard population, as proposed by Pfahl et 
al. (2002) and later revised by BT04 and Liu \& Li (2005). The faint soft  
sources (652) are likely RLOF transients with NS and BH 
accretors in quiescence as proposed by BT04. 
The bright GC sources ($\lesssim 20$; Wang et al. 2002 and MM03) 
can be explained by a population of NS/BH persistent sources 
and transients in outburst (e.g., BT04).
The stellar density of the GC region is higher than that of the solar 
neighbourhood, so it's possible that dynamical interactions may alter 
the number of CVs in the GC.  Tidal captures or exchange interactions may
increase the numbers, but also, the primordial CV population may 
be depleted by tidal disruptions in dense environments (i.e., see Heinke 
et al. 2003).  A full evolutionary model of the GC incorporating binary stellar 
evolution and dynamical interactions should be invoked to gauge these
processes.  Here, we have just demonstrated that field-like (no stellar
interactions) evolution provides enough IPs to explain the GC faint hard X-ray 
population. 

Liu \& Li (2005) conducted a different population synthesis 
study of the GC sources, and it was found that 
most of the {\em Chandra} sources are NS Low Mass X-ray Binary  
transients, with IPs playing a minor role in the Wang et al. and the MM03 
surveys. The IP MT rates obtained by Liu \& Li are much lower ($\sim$ 2
orders of magnitude) than our predictions and those of observational 
estimates. This leads to an underestimate of IP X-ray luminosity in the 
Liu \& Li (2005) 
simulations, and hence a severe underestimate of IP numbers in the GC. 
The Liu \& Li (2005) model is based on the population synthesis code by Hurley, 
Tout \& Pols (2002).  MT is calculated from a formula that does not 
directly incorporate the star's response to mass loss, nor the various
mechanisms for angular momentum loss.  It was originally obtained and
calibrated for binary systems with evolved donors (Algols; Tout \& Eggleton 
1988) and does not appear to work for low mass IP donors.  

We have found that most of the Galactic Centre IPs are either magnetic WDs 
feeding from low-mass MS late-type companions, or double degenerate
systems (e.g., AM CVn). These systems are very unlikely to be detected at
wavelengths other than X-rays due to the high extinction toward the GC,  
and the low luminosity of the IP systems (K-M dwarfs).  
Recently, Laycock et al. (2005) carried out a search for infrared counterparts 
of the GC X-ray sources. It was found that high mass X-ray binaries
(with donors brighter than B2V) are unlikely candidates for the majority of the 
GC faint sources, in agreement with our findings and those of BT04.  
Bandyopadhyay et al. (2005) are conducting a deeper near-infrared
survey which will detect all giant type donors and MS donors with
spectral types earlier than G.  However, if the typical donors in IPs are
indeed K-M MS stars, as found in our study, and also observed for
most Galactic IPs (e.g., M3 V for EX Hya, Dhillon et al. 1997; K5 V for AE Aqr, 
Tanzi et al. 1981), these systems will go undetected in this survey. 

We acknowledge the support of KBN grant 1P03D02228, and thank M.Muno, J.Grindlay, 
K.Belle, X.-W. Liu \& X.-D. Li for helpful discussions.  We also 
thank the anonymous referee for insightful comments.

\begin{deluxetable}{lcccc}
\tablewidth{300pt}
\tablecaption{Galactic Centre Population of IPs}
\tablehead{Model &  WD-MS & WD-giant & WD-WD & Total}
\startdata
Standard CE  & & & &\\
\hspace*{0.5cm}IP$_{\rm frac}$ 1\%  &  694 &  5 &  82 &  781\\
\hspace*{0.5cm}IP$_{\rm frac}$ 5\%  & 3474 & 25 & 410 & 3909\\
\hspace*{0.5cm}IP$_{\rm frac}$ 10\% & 6948 & 50 & 820 & 7818\\
Alternative CE  & & & & \\
\hspace*{0.5cm}IP$_{\rm frac}$ 1\%  &  157 &  1 &  14 &  172\\
\hspace*{0.5cm}IP$_{\rm frac}$ 5\%  &  784 &  5 &  74 &  863\\
\hspace*{0.5cm}IP$_{\rm frac}$ 10\% & 1568 &  9 & 149 & 1726\\
\enddata
\label{tab01}
\end{deluxetable}

\begin{references}
\reference{} Bandyopadhyay, R.M., et al.\ 2005, \mnras, in press (astro-ph/0509346)
\reference{} Belczynski, K., Bulik, T., \& Ruiter, A.J.\ 2005b, \apj, 629, 915 
\reference{} Belczynski, K., Kalogera, V., \& Bulik, T.\ 2002, \apj, 572, 407
\reference{} Belczynski, K., et al.\ 2005a, \apj, submitted (astro-ph/0511811)
\reference{} Belczynski, K., \& Taam, R., 2003 \apj, 616, 1159 (BT04)
\reference{} Belle, K. E., et al.\ 2005, AJ, 129, 1985
\reference{} de Martino, D., et al.\ 2004, Nuclear Physics B, 132, 693
\reference{} Dhillon, V. S., Marsh, T. R., Duck, S. R., \& Rosen, S. R., 1997, \mnras, 285, 95
\reference{} Gansicke, B.T., et al.\ 2005, \mnras, 361, 141
\reference{} Heinke, C. O. et al., 2003, \apj, 598, 501
\reference{} Hurley, J. R., Tout, C. A., \& Pols, O. R., 2002, \mnras, 329, 897
\reference{} Kalogera, V., Webbink, R. F., 1996, \apj, 458, 301 
\reference{} Kroupa, P., Tout, C.A., \& Gilmore, G.,\ 1993, \mnras, 262, 545  
\reference{} Kube, J., Gansicke, B. T., Euchner, F., Hoffmann, B., 2003, \aap, 404, 1159
\reference{} Laycock, S., et al.\ 2005, \apj, submitted (astro-ph/0509783)
\reference{} Launhardt, R., Zylka, R., \& Mezger, P. G., 2002, \aap, 384, 112
\reference{} Liu, X.-W., \& Li, X.-D., 2005, \aap, in press (astro-ph/0512019)
\reference{} Muno, M.P., et al.\  2003, \apj, 589, 225 (MM03)
\reference{} Muno, M. P., et al.\ 2004, \apj, 613, 1179
\reference{} Muno, M. P., Bauer, F. E., Bandyopadhyay, R. M., \& Wang, Q.
             D., submitted (astro-ph/0601627)
\reference{} Nelemans, G., \& Tout, C. A., 2005, \mnras, 356, 753
\reference{} Norton, A. J., \& Watson, M. G., 1989a, \mnras, 237, 853
\reference{} Norton, A. J., \& Watson, M. G., 1989b, \mnras, 237, 715
\reference{} Paczynski, B., 1976, IAUS, 73, 75
\reference{} Patterson, J., 1994, PASP, 106, 209
\reference{} Pfahl, E., Rappaport, S., \& Podsiadlowski, P., 2002, \apj, 571, L37
\reference{} Tanzi, E. G., Chincarini, G., \& Tarenghi, M., 1981, PASP, 93, 68 
\reference{} Tout, C. A., \& Eggleton, P. P., 1988, \apj, 334, 357
\reference{} Tovmassian, G., Zharikov, S., Mennickent, R., \& Greiner, J., 
             2004, ASPC, 315, 15
\reference{} Wang, Q. D., Gotthelf, E. V., \& Lang, C. C., 2002, Nature, 415, 148	
\reference{} Warner, B., 1995, Cataclysmic Variable Stars, Cambridge Astrophysics 
             series, p. 489-501
\reference{} Webbink, R. F., 1984, \apj, 277, 355
\reference{} Wynn, G. A., King, A. R., \& Horne, K., 1997, \mnras, 286, 436

\end{references}
\end{document}